\documentclass[a4paper,11pt]{PoS}

\usepackage{amsmath}
\usepackage{subfigure}
\usepackage{braket}

\newcommand{\be}{\begin{equation}}
\newcommand{\ee}{\end{equation}}
\newcommand{\bea}{\begin{eqnarray}}
\newcommand{\eea}{\end{eqnarray}}

\title{Hypercubic Effects in semileptonic $D \to \pi$ decays on the lattice}

\ShortTitle{Hypercubic Effects in semileptonic $D \to \pi$ decays on the lattice}

\author{V. Lubicz$\,^{(a,b)}$, L. Riggio$\,^{(a)}$, \speaker{G. Salerno}$\,^{(a,b)}$, S. Simula$\,^{(a)}$, C. Tarantino$\,^{(a,b)}$\\

        \textsuperscript{(a)} INFN, Sezione di Roma Tre, Rome, Italy. \\
         E-mail: \email{lorenzo.riggio@gmail.com,} \email{simula@roma3.infn.it}\\
        
        \textsuperscript{(b)} Dipartimento di Matematica e Fisica, Universit\'a di Roma Tre, Rome, Italy. \\
        E-mail: \email{lubicz@fis.uniroma3.it,}  \email{salerno@fis.uniroma3.it,} \email{tarantino@fis.uniroma3.it} \\
        
        \textbf{for the ETM Collaboration}}

\abstract{We present a lattice determination of the vector and scalar form factors of the semileptonic $D \to \pi \ell \nu$ decays, which are relevant for the extraction of the CKM matrix element $\lvert V_{cd} \rvert$ from experimental data. Our analysis is based on the gauge configurations produced by the European Twisted Mass Collaboration with $N_f = 2 + 1 + 1$ flavors of dynamical quarks. We simulated at three different values of the lattice spacing and with pion masses as small as 210 MeV. Quark momenta are injected on the lattice using non-periodic boundary conditions. The matrix elements of both vector and scalar currents are determined for a plenty of kinematical conditions in which parent and child mesons are either moving or at rest. Lorentz symmetry breaking due to hypercubic effects is clearly observed in the data and included in the decomposition of the current matrix elements in terms of additional form factors. Our preliminary estimate for the vector form factor at zero 4-momentum transfer is $f_+^{D \to \pi}(0) = 0.631\,(40)$, which can be compared with the latest FLAG average $f_+^{D \to \pi}(0) = 0.666\,(29)$ available only at $N_f = 2 + 1$.}

\FullConference{34th annual International Symposium on Lattice Field Theory\\
		24-30 July 2016\\
		University of Southampton, UK}

\begin{document}

\section{Introduction and simulation details}

In the Standard Model (SM) weak charged  currents are regulated by the Cabibbo-Kobayashi-Maskawa (CKM) matrix \cite{CKM}. 
An accurate determination of the CKM matrix elements is therefore crucial both for testing the SM and for searching new physics.

In this contribution we present the preliminary results of our determination of the vector and scalar form factors of the semileptonic $D \to \pi \ell \nu$ decays, relevant for the determination of the CKM matrix element $\lvert V_{cd} \rvert$, in the whole kinematical range of values of the squared 4-momentum transfer $q^2$ accessible in the experiments, i.e.~from $q^2 = 0$ to $q^2 = \left( M_D - M_\pi \right)^2$.

We used the gauge ensembles produced by the European Twisted Mass Collaboration (ETMC) with four flavors of dynamical quarks ($N_f=2+1+1$), which include in the sea, besides two light mass-degenerate quarks, also the strange and the charm quarks \cite{Baron:2010bv,Baron:2011sf}.
The simulations were carried out at various lattice volumes and for three values of the lattice spacing in the range $0.06 - 0.09$ fm in order to keep under control the  extrapolation to the continuum limit. 
The simulated pion masses range from $\approx 210$ MeV to $\approx 450$ MeV. 
The gauge fields were simulated using the Iwasaki gluon action \cite{Iwasaki:1985we}, while sea and light valence quarks were implemented with the Wilson Twisted Mass Action \cite{Frezzotti:2003xj,Frezzotti:2003ni}.
The valence charm quark was simulated using the Osterwalder-Seiler action \cite{Osterwalder:1977pc}.
At maximal twist such a setup guarantees the automatic $\mathcal{O}(a)$-improvement \cite{Frezzotti:2003ni,Frezzotti:2004wz}.
More details about the lattice ensembles and the simulation details can be found in Ref.~\cite{Carrasco:2014cwa}.
Gaussian smearing for the pseudoscalar meson interpolating fields has been employed for both the source and the sink.

Momenta were injected on the lattice using non-periodic boundary conditions for the quark fields \cite{Bedaque:2004kc,deDivitiis:2004kq,Guadagnoli:2005be}, obtaining in this way values ranging from $\approx 150$ MeV up to $\approx 650$ MeV.  
The matrix elements of both vector and scalar currents are determined for a plenty of  kinematical conditions in which parent and child mesons are either moving or at rest. 
Lorentz symmetry breaking due to hypercubic effects has been clearly observed in the data (see also Ref.~\cite{Carrasco:2015bhi}). 
In this contribution we present the removal of such hypercubic effects and the determination of the physical, Lorentz-invariant semileptonic vector and scalar form factors.

\section{Lorentz symmetry breaking in the behavior of the scalar and vector form factors}

The matrix elements of the weak vector, $V_\mu = \bar{c} \gamma_\mu d$, and scalar, $V_S =  \bar{c} d$, currents can be extracted from the large time distance behavior of a convenient combination of Euclidean 3-points and 2-points correlation functions in lattice QCD. 
As it is well known, at large time distances 2- and 3-point correlation functions behave as
 \be
      \widetilde{C}_2^M(t) \equiv \left[ C_2^M(t) + \sqrt{C_2^M(t)^2 - C_2^M(T/2)^2} \right] / 2 ~ _{\overrightarrow{t \gg a}} ~ Z_M(\vec{p}_M) e^{-E_M t} / (2 E_M) ~ , 
     \label{C2}
 \ee
 \be
     C^{D \pi}_{\mu,S}(t, t_s) ~ _{\overrightarrow{t \gg a, (t_s - t) \gg a}} ~ \sqrt{Z_D(\vec{p}_D) Z_\pi(\vec{p}_\pi)} ~ \braket{\pi(p_\pi) | V_{\mu,S} | D(p_D)} ~ 
                                            e^{-E_D t - E_\pi (t_s - t)} / (4 E_D E_\pi) ,
     \label{C3}
\ee
where $M$ stands for either the $D$ or the $\pi$ meson, $E_M$ is the meson energy, $t_s$ is the time distance between the source and the sink, ${Z_{\,D}}(\vec{p}_D)$ and ${Z_\pi}(\vec{p}_{\pi})$ are the matrix elements $\lvert \braket{D(\vec{p}_D) \lvert\,\bar{c}\,\gamma_5\,d\,\rvert\,0}\rvert^2$ and $\lvert \braket{\pi(\vec{p}_{\pi})\lvert\,\bar{u}\,\gamma_5\,d\,\rvert\,0}\rvert^2$, where the dependence on the meson momenta $\vec{p}_D$ and $\vec{p}_\pi$ arises from the use of smeared interpolating fields.
In Eq.~(\ref{C3}) the vector and scalar currents are local ones, which renormalize multiplicatively in our setup through the renormalization constants (RCs) $Z_V$ and $Z_P$, respectively. 
As we shall see, the explicit use of the RCs is not required.

The correlation functions can be combined in the five ratios $R_\mu$ ($\mu = 0, 1, 2, 3$) and $R_S$ as
 \be
    R_\mu(t) \equiv 4 p_{D \mu} p_{\pi \mu} \frac{C^{D\pi}_\mu(t, t_s, \vec{p}_D, \vec{p}_\pi) C^{\pi D}_\mu(t, t_s, \vec{p}_\pi, \vec{p}_D)}
        {C^{\pi \pi}_\mu(t, t_s, \vec{p}_\pi, \vec{p}_\pi) C^{DD}_\mu(t, t_s, \vec{p}_D, \vec{p}_D)} ~ _{\overrightarrow{t \gg a}} ~ 
        \left( \braket{\pi(p_\pi) | \hat{V}_\mu |D(p_D)} \right)^2 ,
    \label{Rmu}
 \ee
 \be
    R_S(t) \equiv 4 E_D E_\pi \left( \frac{\mu_c - \mu_\ell}{M_D^2 - M_\pi^2} \right)^2  \frac{C^{D\pi}_S(t, t_s) C^{\pi D}_S(t, t_s)} {\widetilde{C}_2^D(t_s) \widetilde{C}_2^\pi(t_s)} 
        ~ _{\overrightarrow{t \gg a}} ~ \left( \frac{m_c - m_\ell}{M_D^2 - M_\pi^2} \braket{\pi(p_\pi)| \hat{V}_S |D(p_D)} \right)^2 ~ , 
    \label{RS}
 \ee
where $\mu_{\ell(c)}$ are the bare light (charm) quark masses and $m_{\ell(c)}$ are the corresponding renormalized ones. 
In the r.h.s.~of Eqs.~(\ref{Rmu}-\ref{RS}) $\hat{V}_\mu$ and $\hat{V}_S$ are already the renormalized vector and scalar currents, respectively, since the multiplicative RCs $Z_V$ and $Z_P$ cancel out in the ratios (\ref{Rmu}-\ref{RS}).
The matrix elements of the vector and scalar currents can therefore be easily extracted from the plateaux of $R_\mu(t)$ and $R_S(t)$ at large time distances.

Assuming the usual Lorentz-covariant decomposition of the matrix elements of the vector and scalar currents one may have
 \bea
     \label{Vmu_L}
     \braket{\pi(p_\pi) | \hat{V}_\mu |D(p_D)} & = & \left[ P_\mu - (M_D^2 - M_\pi^2) / q^2 \right] f_+(q^2) + q_\mu f_0(q^2) (M_D^2 - M_\pi^2) / q^2 ~ , \\
     \label{S_L}
     \braket{\pi(p_\pi)| \hat{V}_S |D(p_D)} & = & f_0(q^2) (M_D^2 - M_\pi^2) / (m_c - m_\ell) ~ ,
 \eea
where $P \equiv p_D + p_\pi$, $q \equiv p_D - p_\pi$, while $f_{+(0)}(q^2)$ is the vector (scalar) semileptonic form factor.

\begin{figure}[htb!]
\centering
\includegraphics[scale=0.30]{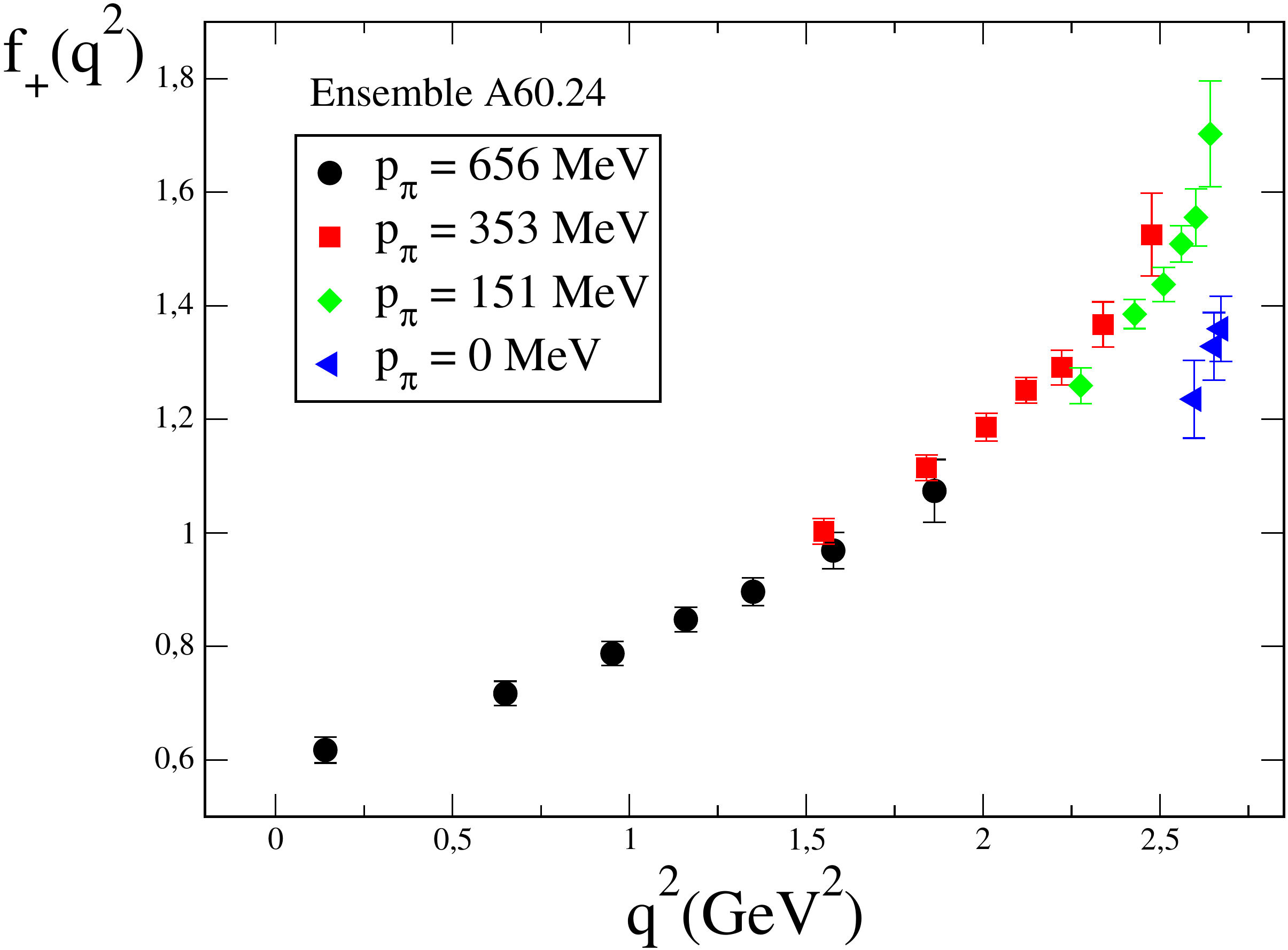}
\includegraphics[scale=0.30]{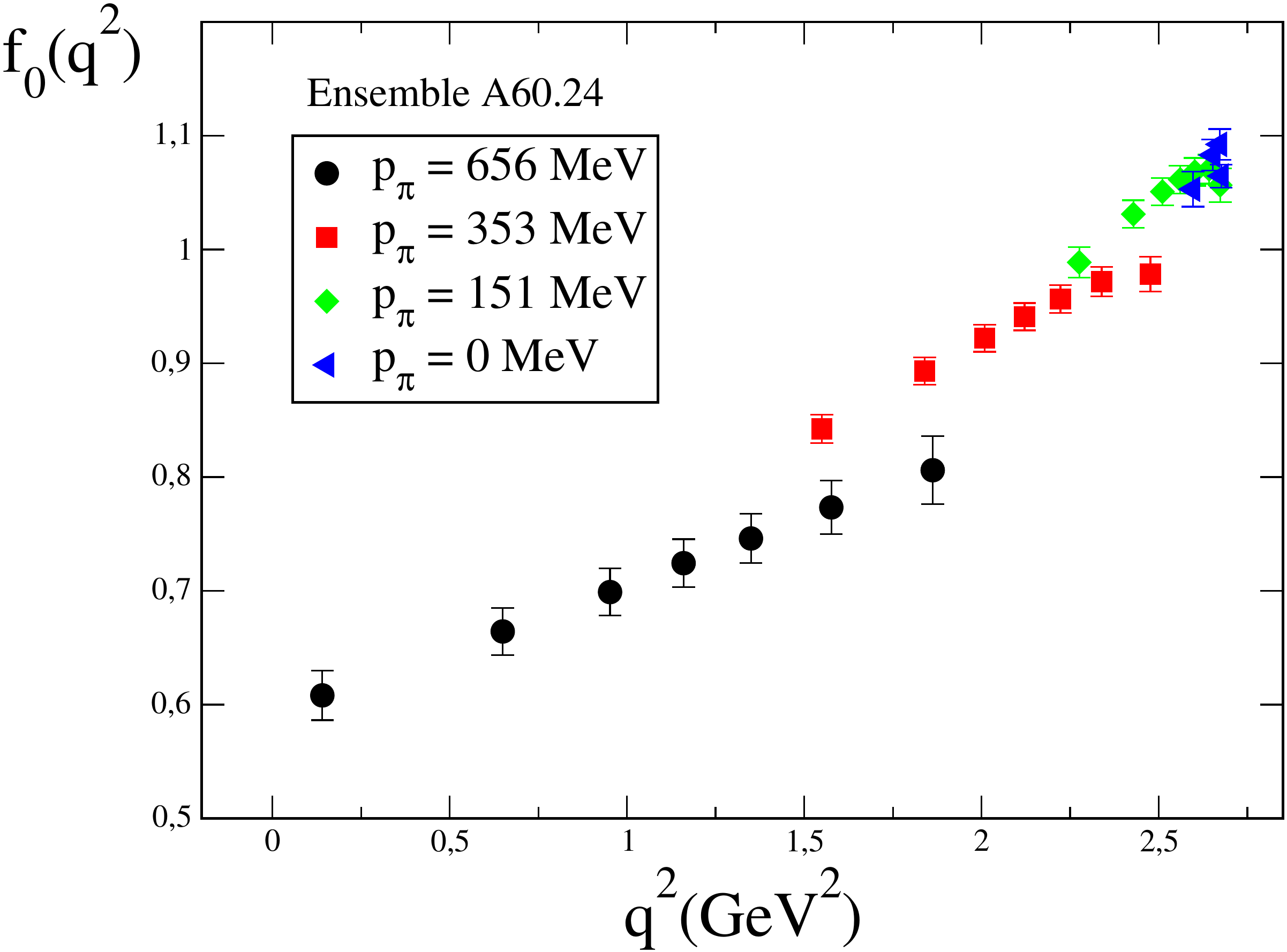}
\caption{\it \footnotesize Momentum dependence of the vector (left panel) and scalar (right panel) form factors in the case of the gauge ensemble A60.24 \protect\cite{Carrasco:2014cwa}. Different markers and colors distinguish different values of the child meson (pion) momentum. The simulated pion mass is $M_\pi \simeq 390$ MeV and the charm quark mass corresponds to its physical value $m_c^{phys}$ from \protect\cite{Carrasco:2014cwa}.}
\label{fishbone}
\end{figure}

For each gauge ensemble and each choice of parent and child meson momenta, $\vec{p}_D$ and $\vec{p}_\pi$, the vector and scalar form factors can be determined as best-fit values of the set of the matrix elements corresponding to the time and spatial components of the vector current and to the scalar current. 
The results, interpolated at the physical value of the charm quark mass $m_c^{phys}$ determined in \cite{Carrasco:2014cwa}, are shown in Fig.~\ref{fishbone} in the case of the gauge ensemble A60.24 (see Ref.~\cite{Carrasco:2014cwa}). 
It can be clearly seen that the extracted form factors are not Lorentz-invariant quantity and a dependence on the value of the child meson momentum is visible.
The investigation and the removal of the observed Lorentz-invariance breaking terms is the subject of the next Section.

\section{Global fit including hypercubic terms}

First we have checked whether the behavior observed in Fig.~\ref{fishbone} may be related to finite volume effects.
Thanks to the presence of two gauge ensembles, A40.24 and A40.32, which share the same pion mass and lattice spacing at different lattice sizes, $L = 24 a$ and $L = 32 a$ (see Ref.~\cite{Carrasco:2014cwa}), it turned out that finite volume effects cannot be the source of the observed behavior of the lattice data. 

Then we tried to describe the breaking of the Lorentz invariance by means of $O(a^2)$ hypercubic effects, since in our setup all the current matrix elements are $O(a)$-improved \cite{Carrasco:2016kpy}. 
In the case of vector current matrix elements possible hypercubic terms have to be odd in the meson momenta and therefore at order $O(a^2)$ we can add to the Lorentz-covariant decomposition (\ref{Vmu_L}) the following hypercubic structure 
 \be
    \braket{\pi(p_\pi) | \hat{V}_\mu^{hyp} |D(p_D)}  = a^2 \left[ q_\mu^3 H_1+ q_\mu^2 P_\mu H_2 + q_\mu P_\mu^2 H_3 + P_\mu^3 H_4 \right] ~ ,
    \label{VH}
 \ee
where $H_i$ ($i=1,...,4$) are additional form factors.
We adopt for them a simple polynomial form in the $z$ variable \cite{Bourrely:2008za}, which depends on both $q^2$ and $m_\ell$, viz.
 \be
    H_i(z) = d_0^i + d_1^i z + d_2^i z^2 ~ ,
    \label{hyp_form_factors}
 \ee
where $d_{0,1,2}^i$ will be treated as free parameters in the fitting procedure .

\begin{figure}[htb!]
\centering
\includegraphics[scale=0.30]{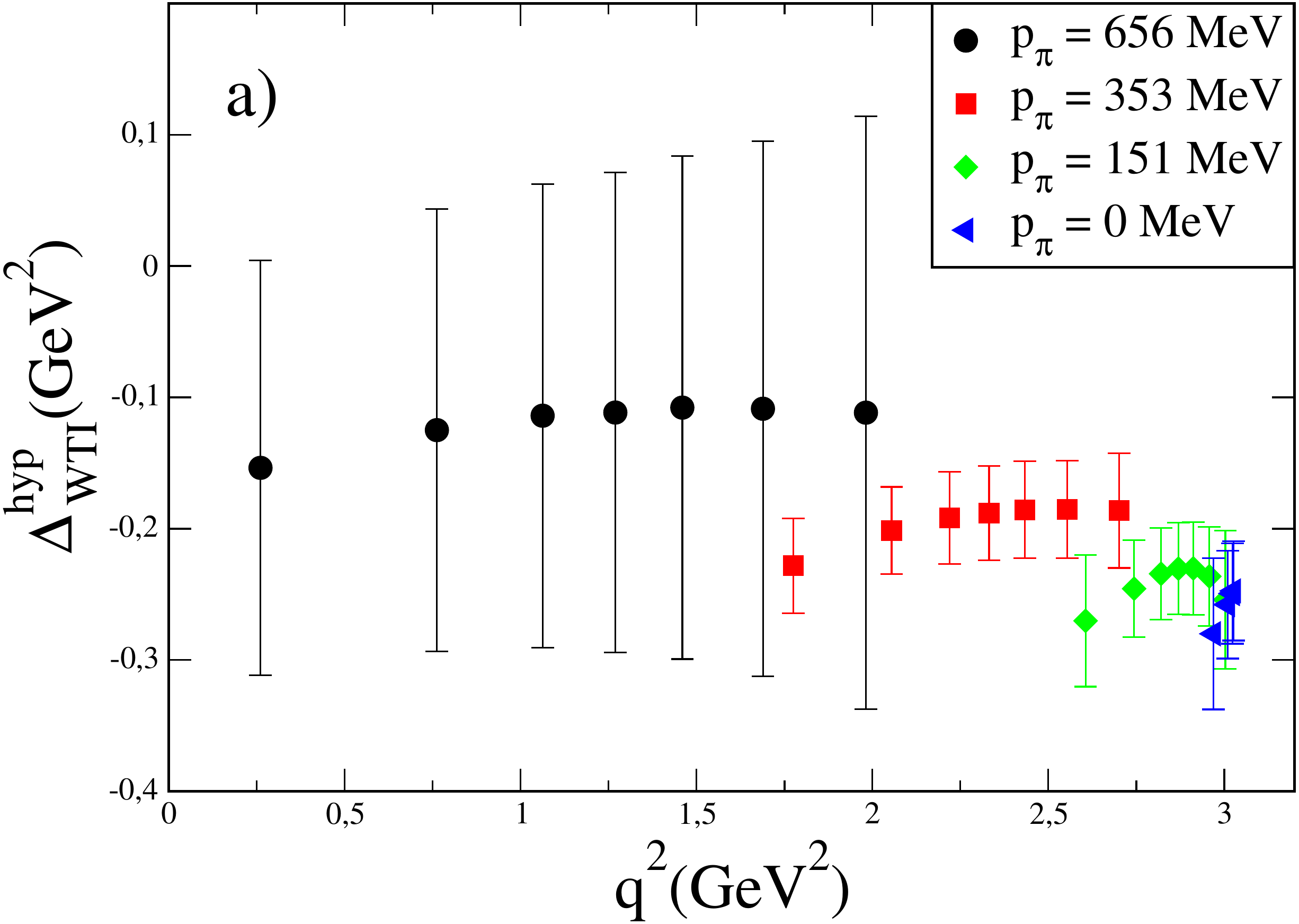}
\includegraphics[scale=0.30]{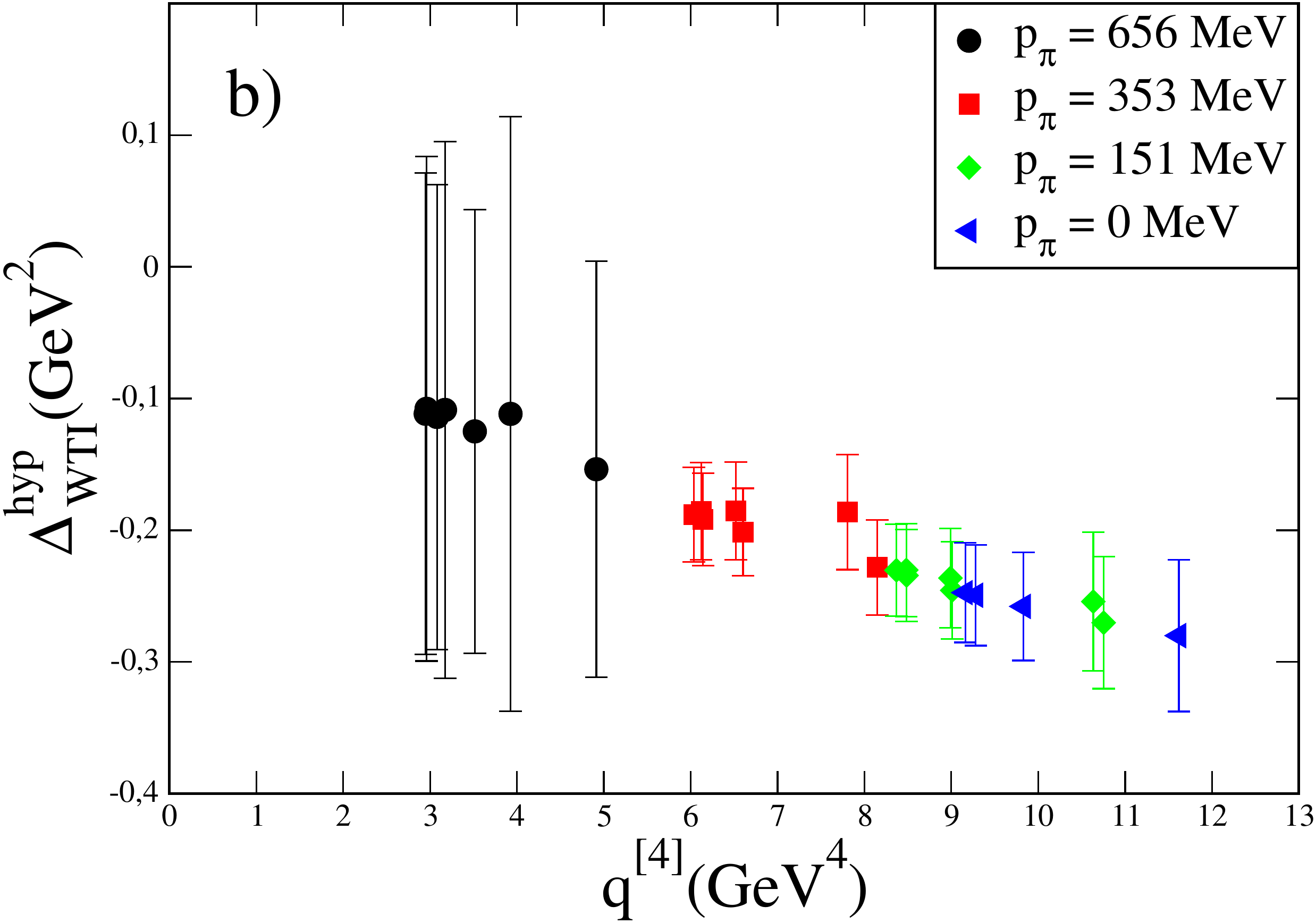}
\caption{\footnotesize \it Results for $\Delta_{WTI}^{\,hyp}$ (see Eq.~(\protect\ref{WI})) versus $q^2$ (a) and $q^{[4]}$ (b) in the case of the gauge ensemble A30.32 \protect\cite{Carrasco:2014cwa}.}
\label{WTI_violation}
\end{figure}

In a similar way one can consider the possible presence of $O(a^2)$ hypercubic terms in the scalar matrix elements.
The Ward-Takahashi identity (WTI), relating the 4-divergence of the vector current to the scalar density, is a good place to look for such hypercubic terms.
We have indeed observed WTI violations that cannot be interpreted as $a^2$ and/or $a^2 q^2$ (Lorentz-invariant) discretization effects.
Thus we have considered the presence of $O(a^2)$ hypercubic effects in the WTI, viz.
 \be
    q^\mu \braket{\pi(p_\pi) | \hat{V}_\mu | D(p_D)} = \left(m_c - m_\ell \right) \braket{\pi(p_\pi) | \hat{V}_S | D(p_D)} + \Delta_{WTI}^{hyp} ~ ,
    \label{WI}
 \ee
where $\Delta_{WTI}^{hyp}$ represent the hypercubic WTI-violating term.
Fig.~\ref{WTI_violation}(a) shows that the WTI-violating term is different from zero and depends on the value of the child meson momentum, in a way similar to that already observed in Fig.~\ref{fishbone}. 
Moreover Fig.~\ref{WTI_violation}(b) suggests a simple linear dependence on the hypercubic invariant $q^{[4]} \equiv \sum_\mu q_\mu^4$, namely
 \be
    \Delta_{WTI}^{hyp} = a^2 ~ q^{[4]} ~ H_5 ~ ,
    \label{WIv}
 \ee
where for $H_5$ we have assumed the simple form $H_5 = d_0^5 + d_1^5 m_\ell$ with $d_{0,1}^5$ being free parameters.

As for the Lorentz-invariant form factors $f_{+,0}(q^2)$ appearing in Eqs.~(\ref{Vmu_L}-\ref{S_L}) we adopt the modified z-expansion of Ref.~\cite{Bourrely:2008za} and impose the condition $f_+(0) = f_0(0) = f(0)$ (as in Ref.~\cite{Carrasco:2016kpy}):
 \be
    f_{+(0)}(q^2) = \frac{f(0) + c_{+(0)} (z - z_0) \left(1 + \frac{z + z_0}{2} \right)}{1 - K_{FSE}^{+(0)} ~ q^2  / M_{V(S)}^2} ~ , 
   \label{z-exp}
 \ee
where $z_0 \equiv z(q^2 = 0$).
Inspired by the hard pion SU(2) ChPT \cite{Bijnens:2010jg}, $f(0)$ can be written as
 \be
    f(0) = b_0 \left[ 1 - \frac{3}{4} \left( 1 + 3 g^2 \right) ~ \xi_\ell \log\xi_\ell + b_1 ~ \xi_\ell + b_2 ~ a^2 \right] 
    \label{ChLim}
 \ee
with $\xi_\ell = 2B m_\ell/ (16\pi^2f^2)$, where $B$ and $f$, determined in \cite{Carrasco:2014cwa}, are the SU(2) low-energy constants entering the LO chiral Lagrangian. 
The coefficients $b_i$ ($i = 0, 1, 2$) are treated as free parameters in the fitting procedure, while $g$ is kept constant at the value $g = 0.61$ \cite{PDG}.
In Eq.~(\ref{z-exp}) the quantity $M_{V(S)}$, representing the vector (scalar) pole mass, is treated as a free parameter, while for the coefficients $c_{+(0)}$ we assume a simple linear dependence on $a^2$.
Finally the comparison of the data in the case of the ensembles A40.24 and A40.32 \cite{Carrasco:2014cwa} suggests the presence of finite size effects (FSEs) in the slope of the form factors. 
We have therefore added the correction $K_{FSE}^{+(0)}$ in the denominator of the r.h.s.~of Eq.~(\ref{z-exp}) to take into account FSEs in a phenomenological way, namely
 \be
    K_{FSE}^{+(0)} = 1 + C_{FSE}^{+(0)} ~ \xi_\ell ~ e^{-M_\pi L} / (M_\pi L) ~ ,
    \label{FSE}
 \ee
where $C_{FSE}^{+(0)}$ are free parameters.

\begin{figure}[htb!]
\centering
\includegraphics[scale=0.30]{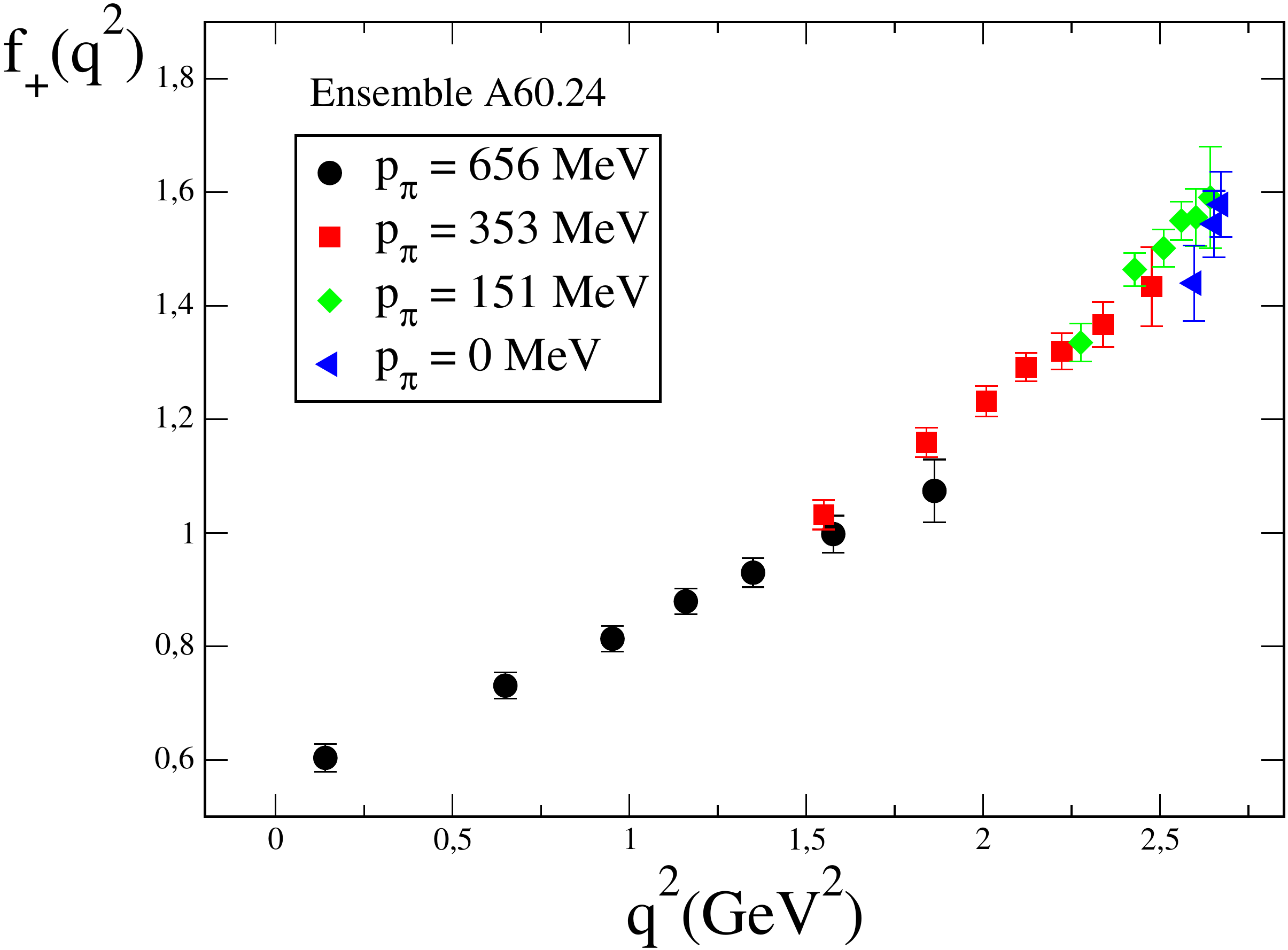}
\includegraphics[scale=0.30]{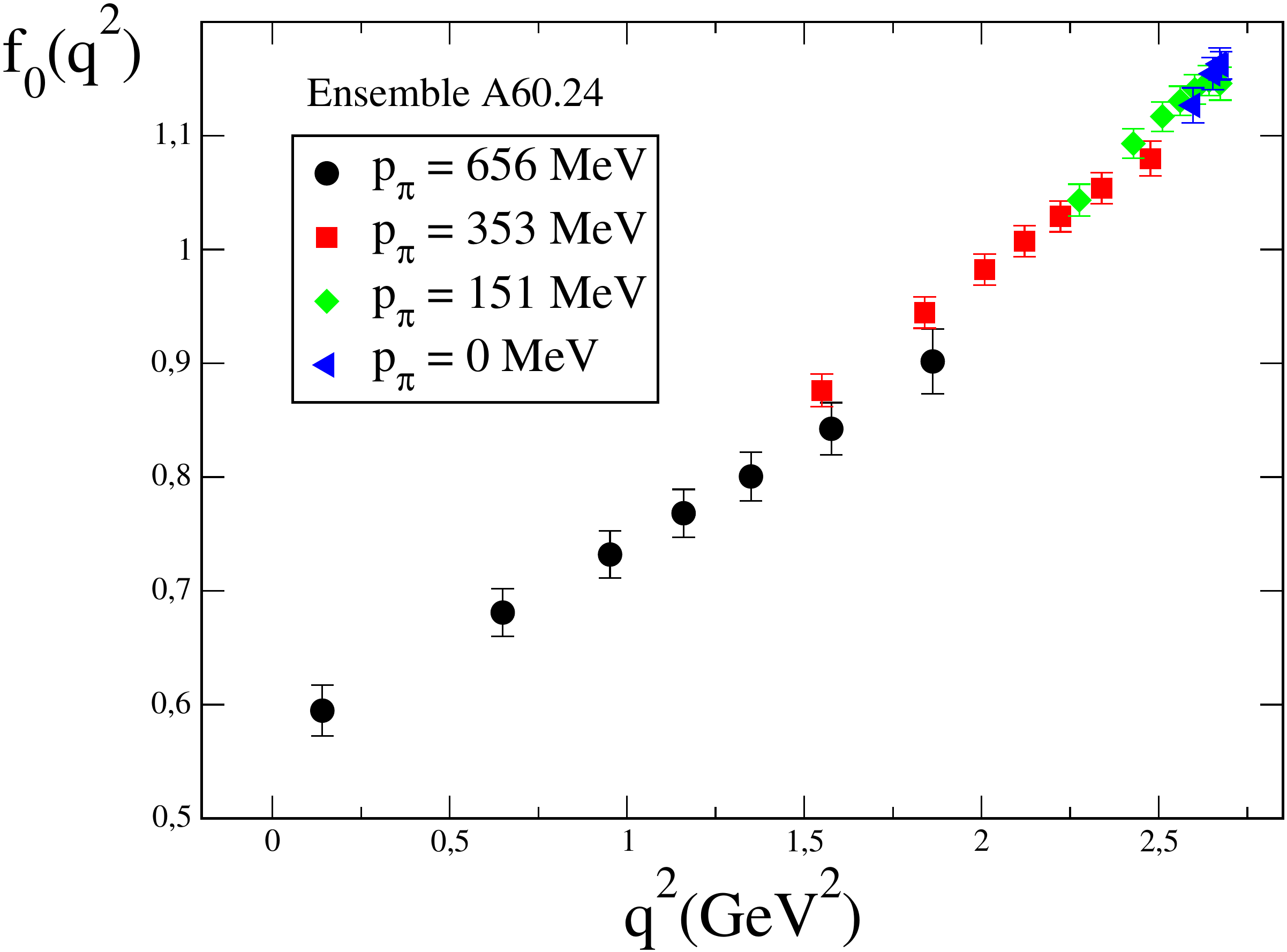}
\caption{\footnotesize \it As in Fig.~\protect\ref{fishbone}, but after the subtraction of the hypercubic effects determined by the global fit.}
\label{corrected}
\end{figure}

Using all the ingredients described above, we have performed a global fit of all the data corresponding to the time and spatial components of the vector current and to the scalar current for all the ETMC gauge ensembles, studying simultaneously the dependence on $q^2$, $m_\ell$ and $a^2$ of the Lorentz-invariant form factors $f_{+,0}(q^2)$ as well as the $q^2$ and $m_\ell$ dependence of the five hypercubic form factors $H_i$ ($i=1,...,5$). 
The quality of the fit is quite good (obtaining a $\chi^2 / \rm{dof}$ equal to $\simeq 1.2$ over more than one thousands lattice points) and will be shown elsewhere. 
Here we limit ourselves to illustrate in Fig.~\ref{corrected} the removal of the hypercubic effects, as determined by the global fitting procedure, in the case of the gauge ensemble A60.24 \cite{Carrasco:2014cwa}.

\begin{figure}[htb!]
\parbox{8.0cm}{~~ \includegraphics[scale=0.30]{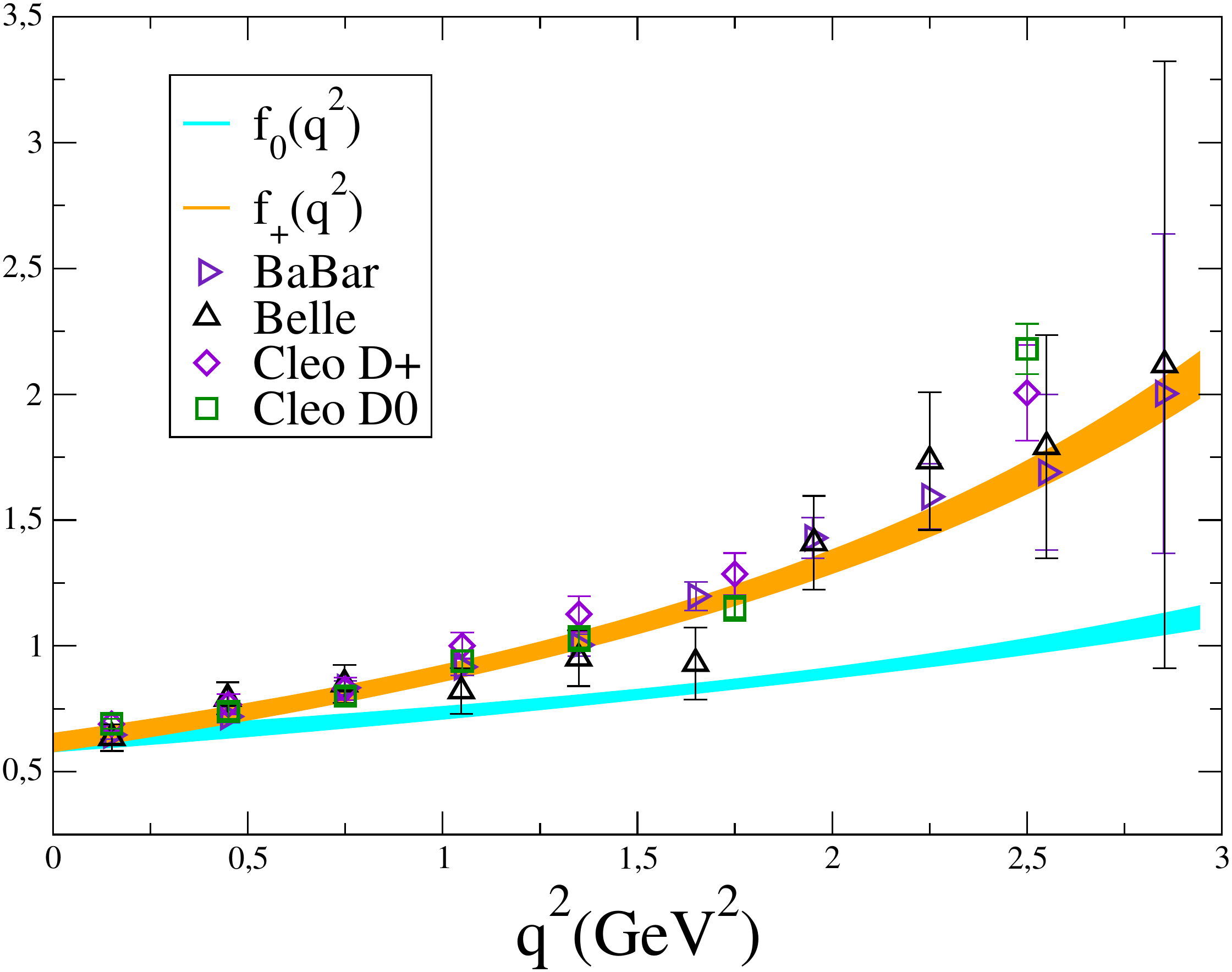}}$~$\parbox{7.0cm}{\caption{\footnotesize \it Results for the vector (orange band) and scalar (cyan band) form factors of the $D \to \pi \ell \nu$ decay extrapolated to the physical point and including the uncertainties related to the statistics, the fitting procedure, the chiral extrapolation and the continuum and infinite volume limits. For comparison the experimental data for $f_+(q^2)$ obtained by Belle \protect\cite{Widhalm:2006wz}, Babar \protect\cite{Lees:2014ihu} and Cleo \protect\cite{Dobbs:2007aa,Besson:2009uv} are shown by the different markers.}
\label{dati_sperimentali}
}
\end{figure}

In Fig.~\ref{dati_sperimentali} the momentum dependence of the vector and scalar form factors $f_+(q^2)$ and $f_0(q^2)$ extrapolated to the physical point is shown in the whole range of values of $q^2$ accessible to the experiments.
In the case of the vector form factor our results are in good agreement with the experimental data obtained by Belle \cite{Widhalm:2006wz}, Babar \cite{Lees:2014ihu} and Cleo \cite{Dobbs:2007aa,Besson:2009uv}.

As for the vector form factor at $q^2=0$ our preliminary results, including the error budget, is
 \be
    f_+^{D \to \pi}(0) = 0.631 ~ (37)_{\rm stat} ~ (14)_{\rm chiral} ~ (08)_{\rm disc} = 0.631 ~ (40) ~ ,
    \label{vec_form_factor}
 \ee
which can be compared with the FLAG average $f_+^{D \to \pi}(0) = 0.666 (29)$ \cite{FLAG}, based on the lattice result obtained at $N_f = 2 + 1$ in Ref.~\cite{Na:2011mc}.

\begin{figure}[htb!]
\centering
\includegraphics[scale=0.30]{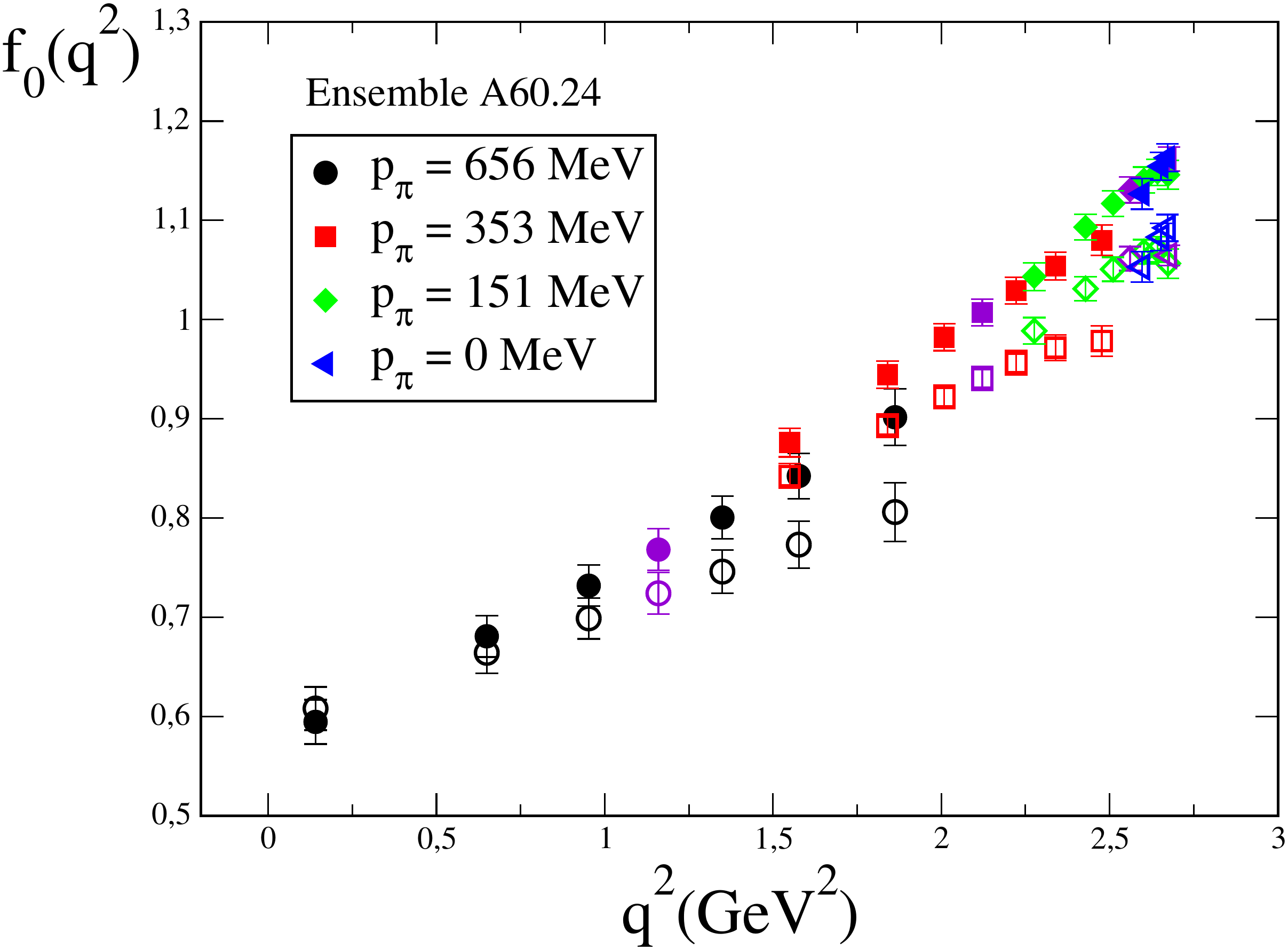}
\includegraphics[scale=0.30]{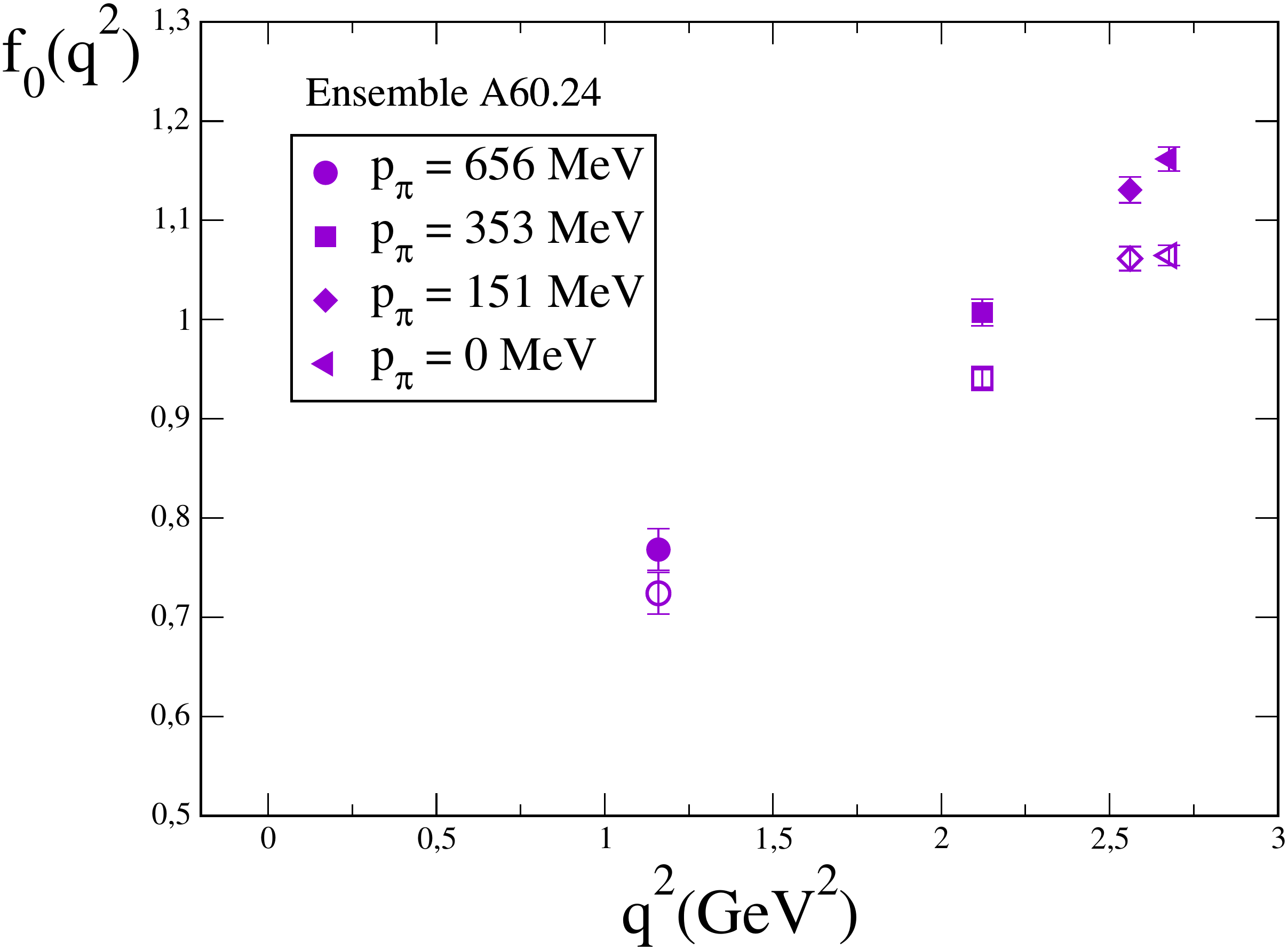}
\caption{\footnotesize \it Behavior of $f_0(q^2)$ for the ensemble A60.24 \protect\cite{Carrasco:2014cwa}. The left panel shows the data corresponding to all the available kinematical configurations, while the right panel only those with the D-meson at rest (purple points). Hollow and filled points represent respectively the data before and after the removal of the hypercubic effects.}
\label{Dfermo}
\end{figure}

The novelty of our analysis with respect to previous studies of the semileptonic $D \to \pi $ form factors is the use of a plenty of kinematical configurations corresponding to parent and child mesons either moving or at rest. 
We notice that using only a limited number of kinematical conditions, for instance the Breit-frame ($\vec{p}_D = - \vec{p}_\pi$) or the D-meson at rest, the presence of the hypercubic effects may not be detected.
This is nicely illustrated in Fig.~\ref{Dfermo}, which shows our data of the scalar form factor $f_0(q^2)$ for the whole set of kinematical configurations and for the case of the D-meson at rest. 
The limited set of data with $\vec{p}_D = 0$ does not show any evidence of the Lorentz-symmetry breaking, although its presence is quite visible by comparing the data before and after the removal of the hypercubic effects determined in the global fitting procedure. 
This may lead to a systematic error in the extraction of the momentum dependence of the physical form factors.

\section{Acknowledgements}
\footnotesize{We acknowledge the CPU time provided by PRACE under the project PRA067 on the BG/Q systems Juqueen at JSC (Germany) and Fermi at CINECA (Italy), and by the agreement between INFN and CINECA under the initiative INFN-LQCD123. L. R. thanks INFN for the support under the SUMA computing project (https://web2.infn.it/SUMA).}

\end{document}